# Whither does the Sun rove?

Alejandro Gangui, IAFE/Conicet and University of Buenos Aires, Argentina

*If one asked some friends where on the horizon they should expect to see the sunrise, half of the answers would be "in the east". Of course, something analogous would happen with the sunset and the west. However, sunrise and sunset virtually never occur at these cardinal points. In fact, those answers correctly describe observations only during the equinoxes, when either autumn or spring begin. Once we recall this, the next natural question to ask ourselves is: how far from the east (or from the west) the rising (or setting) Sun is located for a given latitude of the observer and for a given day of the year. In this paper we supply some simple tools to easily visualize the angular (southward or northward) departure of the rising and setting Sun on the horizon from the east-west direction in a pictorial way, without the need of mathematics. These tools have proven a valuable resource in teaching introductory physics and astronomy courses.*

It is an observational fact that the Sun does not rise or set exactly at the same time in every month of the year. In fact, there are notable changes even from one week to the next. Although the Sun at solar midday is high in the sky during the summer, and low above the horizon during the winter, the "circular arc" trajectory that it describes during the day is just a manifestation of our planet's own rotation. As a consequence, the solar arcs on the sky (a slightly different one for each day of the year) form planes orthogonal to the axis of rotation of the Earth.

The celestial equator, the projection of the Earth's equator into space, is orthogonal to the Earth's axis (see Figure 1). Hence, the arc of the Sun in the sky is always parallel to the celestial equator.

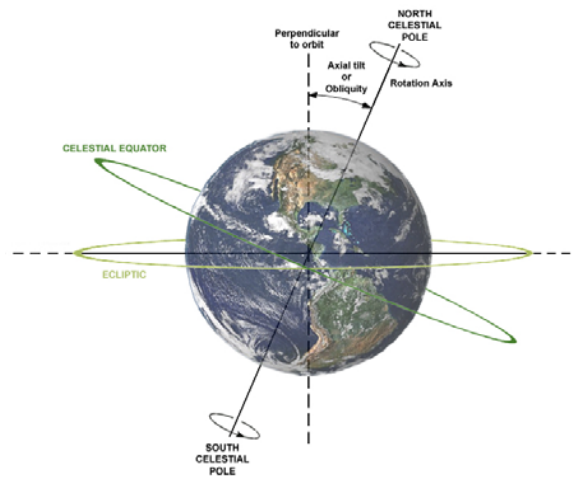

*Figure 1: The Earth as viewed from an external observer on the ecliptic. The figure indicates the orientation of celestial equator. The axis of rotation of the Earth points, on the north (south), towards the north (south) celestial pole. This axis is tilted 23.5° with respect to the direction perpendicular to the plane of the orbit.*

Imagine we are located on the Sun and we look toward the Earth, as in Figure 1 (the Sun is located out of the picture). Our line of sight is parallel to the ecliptic. Imagine now that a friend is located on the equator. What would the arc of the Sun look like in her sky? Let the Earth spin and she will see the Sun appearing on the eastern horizon, rising perpendicularly to it. The Sun will then make a circular circuit roughly above her head to finally set perpendicularly to her western horizon. Figure 2 shows the view corresponding to her position, both in perspective (a) and from the west (b).

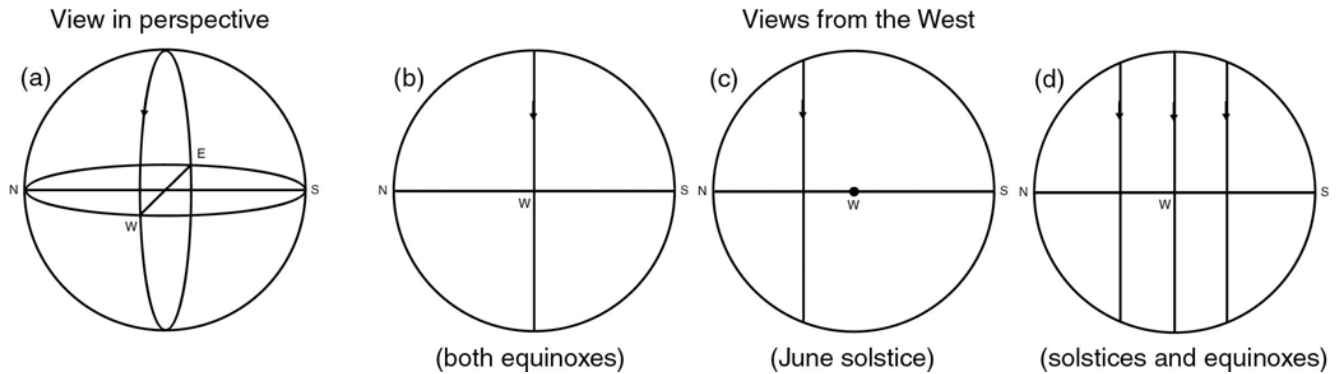

*Figure 2: The daily trajectory of the Sun corresponding to an observer located on the equator, viewed from the "outside" of the celestial sphere: (a) depicts a view in perspective, with the plane of the horizon and the four cardinal points; (b) shows a side view during both equinoxes; (c) shows the same view during the solstice of June. Finally, (d) shows the four daily trajectories of the Sun corresponding to both solstices and both equinoxes. Note that on this scale, the size of the Earth is negligible compared with the celestial sphere (the larger circle in the pictures).*

Now imagine that roughly three months have passed, the Earth has completed a quarter of its orbit, and that we are still on the Sun looking toward the Earth. The situation is again represented in Figure 1, but now we (and the Sun) are on the right-hand side of the picture. The northern hemisphere is tilted toward the Sun. It is the June solstice (summer solstice for the northern hemisphere). Now let the Earth spin. During the day the Sun passes through the zenith for observers located at latitude +23.5°. In fact, that day the angular distance of the Sun from the celestial equator (what we call the "declination" of the Sun) is precisely +23.5°. But for our friend located on the equator (at latitude 0°) the situation is rather different. She will see that the Sun is shifted always toward the north, from sunrise until sunset. The situation is depicted in Figure 2 (c).

If we let the Earth go around the Sun another half of its orbit (roughly, a six-month period), we reach the December solstice. Now the Sun is on the left-hand side of the Earth in Figure 1, and our friend on the equator will always see the Sun shifted toward the south (a mirror image of the previous situation). Figure 2 (d) shows the three (actually four) trajectories of the Sun during the day for these particular situations, corresponding to an observer on the equator. There are three lines representing four trajectories because both equinoxes are given by the same central vertical line. All 363 (=365-2) daily trajectories of the Sun are similar vertical lines lying in between the solsticial ones. Figure 3 shows the maximum declination of the Sun: +23.5° (or -23.5°), toward the north (or the south) of the celestial equator. This fixes the separation of the three vertical lines in Figure 2 (d).

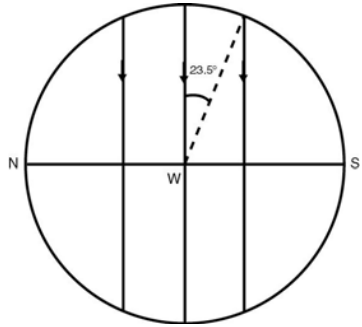

*Figure 3: The angular position of the Sun with respect to the celestial equator is measured in degrees and is given by the declination. In the figure we show two extreme declinations for the Sun: +23.5°, when it is located at its maximum angle towards the north (vertical line on the left) and -23.5°, when it reaches its maximum angle towards the south (vertical line on the right, with the angle shown in the figure). The Sun's declination covers roughly 47° (= 2 x 23.5°) in roughly a six-month period (although not at a constant speed).*

Away from the equator

Now let our friend move on from the equator. As she travels, her latitude increases: to positive "N" values above (north of) the equator, or to positive "S" values below (south of) it. As she travels, also her local horizon will gradually bend with respect to the diurnal trajectories of the Sun of Figure 3. We draw this change in Figure 4, where we show the situation corresponding to four different locations: latitudes 35° and 60° north and south of the equator. Note that the angle 23.5°, corresponding to the maximum declination of the Sun, is fixed. Note also that the trajectories of the Sun get tilted with respect to the vertical of the site: more tilted for higher latitudes.

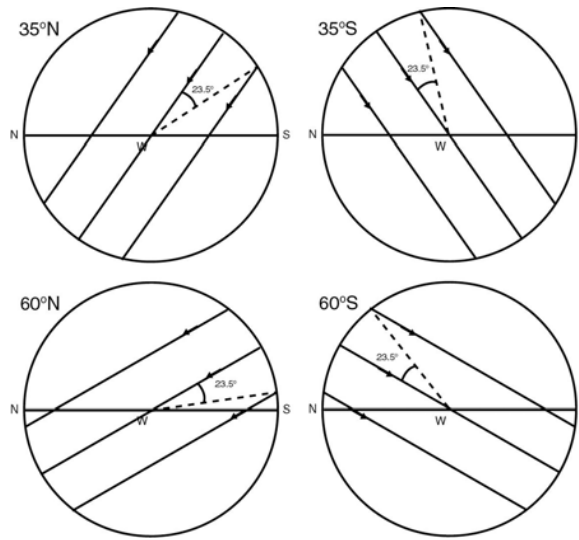

*Figure 4: The daily trajectories of the Sun corresponding to observers at four different latitudes, both north and south of the equator (see main text for details).*

Furthermore, from Figure 3 (observer at the equator) we see that during the solstices both sunrise and sunset are located 23.5° away from the east-west direction. So, as we mentioned, claiming that the Sun always rises in the east (and sets in the west) is clearly a mistake. But from Figure 4 we see that this claim is more of a mistake the higher the latitude of the observer. For latitude 60° north, for example, the rising or setting points of the Sun's circuit are closer to the north cardinal point (in the June solstice) or to the south cardinal point (in the December solstice) than they are to the east or to the west.

A simple tool

Of course, one should be careful with these diagrams, as the real angles that we measure live on the celestial sphere and not on the Euclidean plane (recall that the sum of the interior angles of a triangle drawn on the sky exceeds 180°). Although one can perform some graphical [1] and analytical [2] computations, in order to estimate the angle formed between the sunrise location and the east, for every day of the year (that is, for every possible declination of the Sun), the need of spherical geometry complicates the analysis.

Our aim here is not to pursue this task, but rather to supply a useful and simple tool that will help students visualize the angular (southward or northward) departure of the rising and setting Sun from the east-west direction in a pictorial way. In Figures 5 and 6 we show two diagrams, which should be employed together. Just make a copy of both of them, print the one of Figure 6 in a transparency and put it centered over the diagram of Figure 5. The arrow in Figure 6 is used to indicate the latitude of the observer, as if it was a "moving hand" on the angular "dial" (expressed in degrees) of Figure 5. This allows us to choose any latitude we wish (some particular latitudes are already written down in the dial). Given our previous explanations for Figures 1 to 4, the *user guide* for this new tool should now be transparent enough.

This "sunrise-sunset dial" also allows us to reproduce many other situations, some of which are responsible for headaches suffered by both physics and astronomy teachers. For instance, let our friend be located in one of the tropics (the Tropic of Cancer, latitude 23.5°N, in the northern hemisphere or the Tropic of Capricorn, 23.5°S, in the southern one). Using the dial we can check that, on the day of the summer solstice, the trajectory of the Sun culminates in the zenith: an observer in the tropic has just one day like this. Observers in lower latitudes will find the Sun in their vertical line exactly twice per year (once in spring and another time in summer), whereas observers in higher latitudes will never find the Sun in their zenith (this is the situation depicted in Figure 4).

Let our friend now be located either in the Arctic or in the Antarctic Circle, latitudes 66.5° north or south, just bordering the polar regions. Employing our instrument we easily see that during the winter solstice, the trajectory of the Sun crosses the dial at positions south or north, respectively: both the sunrise and the sunset take place at the very same point on the horizon, namely, exactly at the south or at the north cardinal point, respectively. Clearly, for higher latitude regions, the Sun will not rise at all: this is the *polar night* (when the night lasts for more than 24 hours). On the other hand, in the opposite polar region, observers will enjoy the *midnight sun* (when the Sun stays above the horizon for more than 24 hours): we clearly see this with our dial, as the trajectory of the Sun lies always above the horizon.

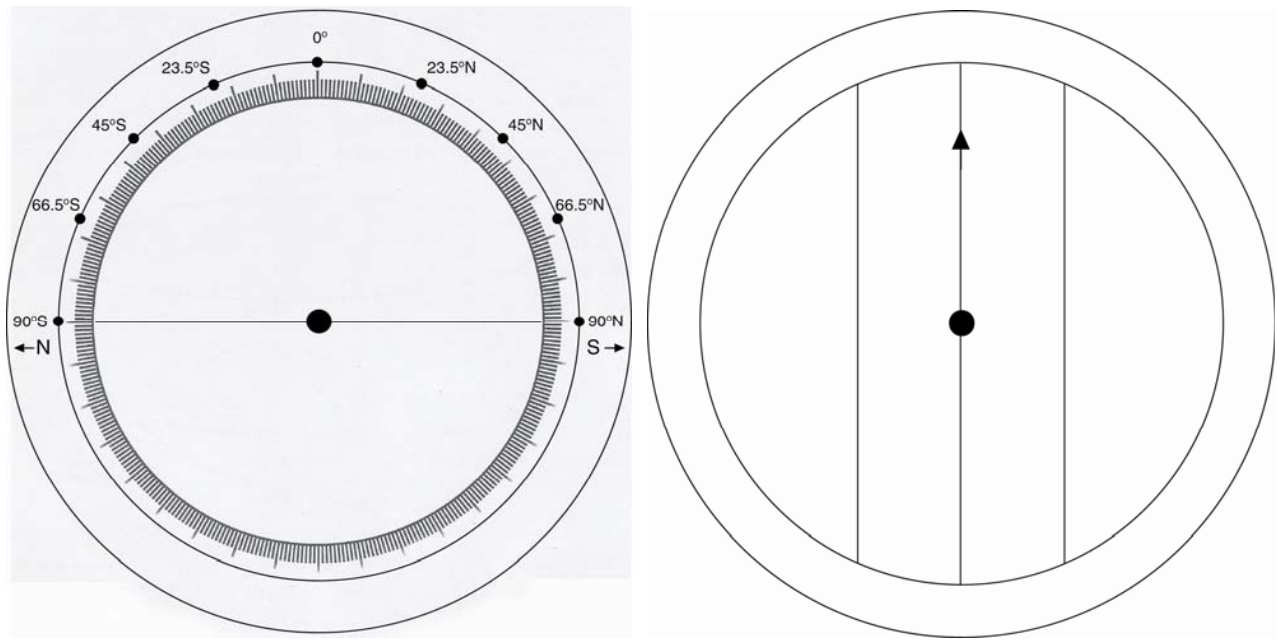

*Figures 5 (left panel) and 6 (right panel): The arrow of Figure 6 should point to the observer's latitude as recorded in Figure 5. The straight line containing this arrow corresponds to the trajectory followed by the Sun during the days of the Equinoxes. The other two straight lines mark this trajectory during the solstices: the one on the left always corresponds to the solstice of June; the one on the right, to the solstice of December [3].*

The higher the latitude, the more the number of polar nights the observer will endure, and the more the number of days with a midnight sun her "antipodes" will enjoy (we can readily see this by moving the hand of our dial to higher latitudes). Reaching the extreme situation in which our friend is located exactly at one of the poles, latitude 90°, we see that the arrow of our instrument will point horizontally: during six months the Sun will not rise, while in the other six months it will not set. She will either have six months of daylight or six months of "night". Of course, due to the fact that the Sun is not point-like, these numbers change a little bit (when the Sun is partially above the horizon one should count it as "daytime"). Also, due to refraction in the Earth's atmosphere, many of these polar "nights" are not dark at all.

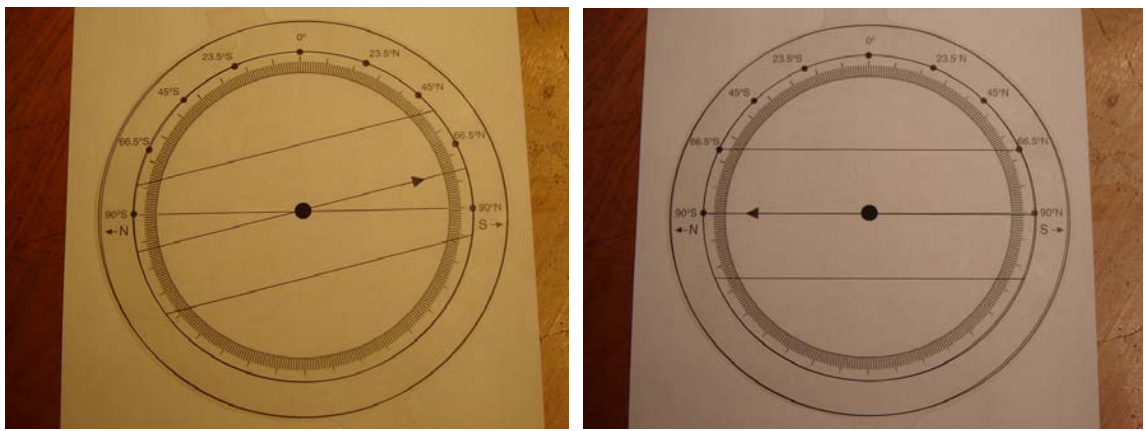

*Figure 7: The simple instrument constructed and employed by the author for his physics and astronomy courses, shown here "dialed" for two different latitudes: one within the Arctic Circle (left panel), the other one corresponding to an observer located exactly on the South Pole (right panel).*

Final remarks

We have discussed at length some issues related to the actual location on the horizon where the Sun rises and sets. We recalled that on just two days in the year the Sun rises in the east and sets in the west. During these equinoctial days, day and night last the same amount of hours (apart from details due to refraction and due to the finite size of the Sun's disc). We have seen that the rest of the year, the Sun does not "touch" the east or west cardinal points when it crosses the horizon.
In this paper, we introduced a simple tool that allows students to quickly understand and estimate qualitatively how far away from the east-west line the Sun rises or sets, regardless of the latitude of the observer and of the chosen day of the year. This tool, and specially its limitations, serves also as a motivation to discuss the relevance of spherical geometry when one wants to make precise computations involving distances and angles in the celestial sphere. This will hopefully lead to further studies more compelling to the students' mind, therefore improving their attitude toward mathematics and science in general.

*Alejandro Gangui is staff researcher at the Institute for Astronomy and Space Physics (IAFE) and is professor of physics at the University of Buenos Aires (UBA). His personal website is http://cms.iafe.uba.ar/gangui/ .*

This article was published in
The Physics Teacher, vol 49, February 2011, pp. 91-93.